\documentclass[aps,prb,reprint,groupedaddress]{revtex4-1}

\usepackage{graphicx}

\usepackage{bm}

\usepackage{hyperref}

\usepackage{float}
\usepackage{amsmath}

\begin{document}

\title{Hybrid surface waves in two-dimensional Rashba-Dresselhaus materials}

\author{Dmitry Yudin}
\affiliation{ITMO University, Saint Petersburg 197101, Russia}
\affiliation{Division of Physics and Applied Physics, Nanyang Technological University 637371, Singapore}

\author{Dmitry R. Gulevich}
\affiliation{ITMO University, Saint Petersburg 197101, Russia}

\author{Ivan A. Shelykh}
\affiliation{ITMO University, Saint Petersburg 197101, Russia}
\affiliation{Division of Physics and Applied Physics, Nanyang Technological University 637371, Singapore}
\affiliation{Science Institute, University of Iceland IS-107, Reykjavik, Iceland}

\date{\today}

\begin{abstract}
We address the electromagnetic properties of two-dimensional electron gas confined by a dielectric environment in the presence of both Rashba and Dresselhaus spin-orbit interactions. It is demonstrated that off-diagonal components of the conductivity tensor resulting from a delicate interplay between Rashba and Dresselhaus couplings lead to the hybridization of transverse electric and transverse magnetic surface electromagnetic modes localized at the interface. We show that the characteristics of these hybrid surface waves can be controlled by additional intense external off-resonant coherent pumping.
\end{abstract}

\maketitle

\section{Introduction}

Tremendous progress in laser physics and nanotechnology has stimulated activity in the field of light-matter interactions in low-dimensional systems. A plethora of light-induced phenomena ranging from pump-probe spectroscopy \cite{Kling2008} to ultrafast magnetization dynamics \cite{Kirilyuk2010} form a basis of our understanding. In most cases, light-matter coupling is considered to be small enough so that all physically relevant effects, associated with the transition from one eigenstate to another and accompanied by absorption or emission of light quanta, can be captured perturbatively. However, this simple picture breaks when the light-matter interaction is intensive enough and a regime of strong light-matter interaction is achieved. It was recently predicted that in this regime, the transport and optical properties of low-dimensional electronic structures can change dramatically~\cite{Morimoto2009,Wittmann2010,Tarasenko2011,Pervishko2015,Morina2015}. In particular, it can be expected that external pumping may be used as a unique tool to address surface electromagnetic waves in low-dimensional guiding structures~\cite{Polo2011}. These waves can be subdivided to belong to one of two classes: surface electromagnetic waves residing at the boundary between two media with permittivities of opposite signs, and surface electromagnetic waves emerging due to the optical anisotropy of the surrounding media with permittivities of the same sign (Dyakonov waves \cite{Dyakonov1988}). The simplest example of the waves of the first type is a surface plasmon polariton at a metal-dielectric interface \cite{Schuller2010,Pitarke2007}. Plasmon excitations are widely used in surface spectroscopy and have been demonstrated to be capable of supersensitive sensoring \cite{Zalyubovskiy2012}, and can be used for local spectroscopy and nanolithography with unprecedented spatial resolution which dramatically exceeds the Rayleigh limit \cite{Lozovik1999}. Another possible application of surface plasmons and plasmon polaritons is in the field of ultrafast information transfer, which is known to be faster than that with electron current pulses. The wide range of the applications of plasmonics has allowed it to remain among the trends of modern applied science. It should be noted, however, that all these promising applications are significantly restricted by the strong damping of plasmons, and the search for novel systems with low damping rates, e.g., based on doped graphene \cite{Kumar2016}, becomes an actual task.

\begin{figure*}
\centering
\includegraphics[width=\linewidth]{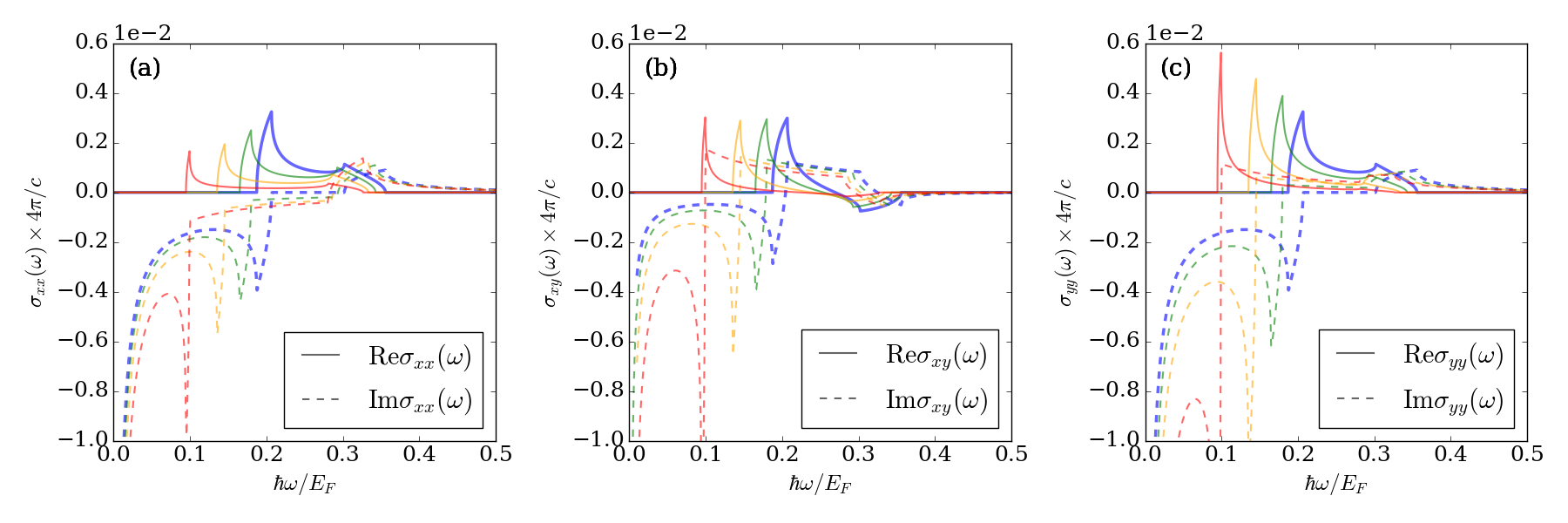}
\caption{Numerical results uncovering the behavior of the diagonal and off-diagonal components of the conductivity tensor: (a) $\sigma_{xx}(\omega)$, (b) $\sigma_{xy}(\omega)=\sigma_{yx}(\omega)$, and (c) $\sigma_{yy}(\omega)$. The real part of the conductivity is finite within a certain frequency range, stemming from the trade-off between Rashba and Dresselhaus couplings, while the use of off-resonant pumping allows for further manipulation with this interval. Imaginary parts are sign alternating with respect to the frequency of a probing field $\omega$. The nonzero $\sigma_{xy}(\omega)$ guarantees hybridization between the TE and TM modes to form hybrid surface waves. The thick blue solid line corresponds to the case of the absence of external pumping. The cases for off-resonant pumping with light-matter coupling strengths $\gamma=0.4,0.6$, and $0.8$ are presented by the green, orange, and red lines, respectively.}
\label{fig:numerics}
\end{figure*}

One of the possible ways to reduce damping is to use a conducting interface sandwiched between two conventional insulators instead of a metal-dielectric boundary. In this geometry, a surface electromagnetic mode is formed, because the tangential component of the electric field at the conducting interface generates surface current density which leads to a discontinuity in the tangential magnetic field \cite{Slepyan1999, Mehrany2004, Mikhailov2007}. Examples of  such a conducting interface are two-dimensional electron gas (2DEG) in quantum-well structures or electrons in graphene characterized by the linear dispersion relation \cite{Iorsh2013,Yudin2015,Kumar2016}. In this paper we focus on a theoretical analysis of the optical response and propagation properties of hybrid surface electromagnetic waves in thin film semiconductors with a spin-orbit interaction (SOI) of the Rashba and Dresselhaus type. The competition between these couplings was shown to lead to such phenomena as the emergence of Hall-type conductivity \cite{Bryksin2006,Yudin2016} and anisotropic plasmon dynamics \cite{Badalyan2009,Cruz2014}, which makes the dynamics of Rashba-Dresselhaus systems more rich compared with the systems with Dresselhaus or Rashba SOIs only \cite{Pletyukhov2006}. The rest of the paper is organized as follows: In Sec.~\ref{sec:conductivity}, based on the Kubo formula we work out the components of the conductivity tensor. We show that the presence of both Rashba and Dresselhaus SOIs results in a finite value of the off-diagonal components of the conductivity tensor in the whole spectral range. In what follows, we demonstrate that the latter leads to the hybridization of surface electromagnetic waves in Sec.~\ref{sec:waves}, and elaborate on the role of external pumping on their properties in Sec.~\ref{sec:pump}. Finally, we summarize our main findings and provide a short outlook in Sec.~\ref{sec:conclusions}.

\section{Optical conductivity}\label{sec:conductivity}

The single-particle Hamiltonian of spin-orbit-coupled 2DEG can be expressed as a sum of the terms corresponding to the kinetic energy and SOI, $\hat H=p^2/(2m)+\hat H_\mathrm{SO}$. In the case of a quantum well of a zinc-blende structure grown in the [001] direction, the SOI associated with the momentum-linear terms reads

\begin{equation}\label{so}
\hat H_\mathrm{SO}=\left(\alpha p_y+\beta p_x\right)\hat\sigma_x-\left(\alpha p_x+\beta p_y\right)\hat\sigma_y,
\end{equation}

\noindent and is characterized by two real parameters, the Rashba spin-orbit-coupling strength $\alpha$, resulting from the asymmetry of the confining potential, and the Dresselhaus spin-orbit-coupling strength $\beta$, stemming from the lack of inversion symmetry of a crystal; $\hat\sigma_i$ are Pauli matrices and $\mathbf{p}=\left(p_x,p_y\right)$ is the electron momentum restricted to 2D. Following the standard paradigm, the charge current  conductivity tensor is estimated as a linear response to a frequency-dependent and spatially homogeneous weak electric field. The angular anisotropy of the energy splitting due to the simultaneous presence of Rashba and Dresselhaus interactions gives rise to a finite-frequency response with spectral features significantly different from those of a pure Rashba or Dresselhaus model. 
The effect of linear Dresselhaus coupling was shown to have a profound impact on the transport properties of the systems \cite{Miller2003,Schliemann2003a,Schliemann2003b,Mishchenko2003,Ganichev2004,Sinitsyn2004,Shen2004,Erlingsson2005}. For a probing field with frequency $\omega$ the dynamical conductivity  is given by

\begin{equation}
\sigma_{ab}(\omega)=\frac{1}{\hbar\omega}\int\limits_0^\infty dt\langle[\hat{j}_a(t),\hat{j}_b(0)]\rangle e^{i(\omega+i\delta)t},
\label{kubo}
\end{equation}
where $\hat{\mathbf{j}}=-e\nabla_\mathbf{p}\hat H$ is the current operator. The angular brackets in Eq.~(\ref{kubo}) denote quantum and thermal averaging, while a positive infinitesimally small $\delta>0$ guarantees the convergence in the upper limit. Subscripts $a,b=x,y$ correspond to Cartesian components of the current operator $\hat{\mathbf{j}}$. To simplify the calculations we worked out expression (\ref{kubo}) in the limit of vanishing temperature $T=0$ and in the absence of disorder (see Fig.~\ref{fig:numerics}). When evaluating the real part of the conductivity tensor for a spin-orbit-coupled material, there arise several integration areas separated by the boundary frequencies $\omega_-<\omega_a<\omega_b<\omega_+$ (see Refs.~\cite{Maytorena2006,Yudin2016}).
A delicate interplay between Rashba and Dresselhaus couplings results in the real part of the conductivity tensor being nonzero in a broader region compared to the pure Rashba or Dresselhaus model, limited by $\omega_-$ and $\omega_+$.
Meanwhile, its imaginary part is negative in a certain range of the frequencies of the probing field, which may potentially lead to the emergence of transverse electric (TE) surface modes~\cite{Falko1989,Mikhailov2007}. Hence, the most important modification arising due to the competition between two different linear spin-orbit interactions in comparison with a pure Rashba or Dresselhaus system or 2DEG stems from the new boundaries of the electron-hole continuum, corresponding to the Landau damping region characterized by $\mathrm{Re}\,\sigma_{ab}(\omega)\neq0$. 

\begin{figure*}[htp!]
\begin{center}
\includegraphics[width=\linewidth]{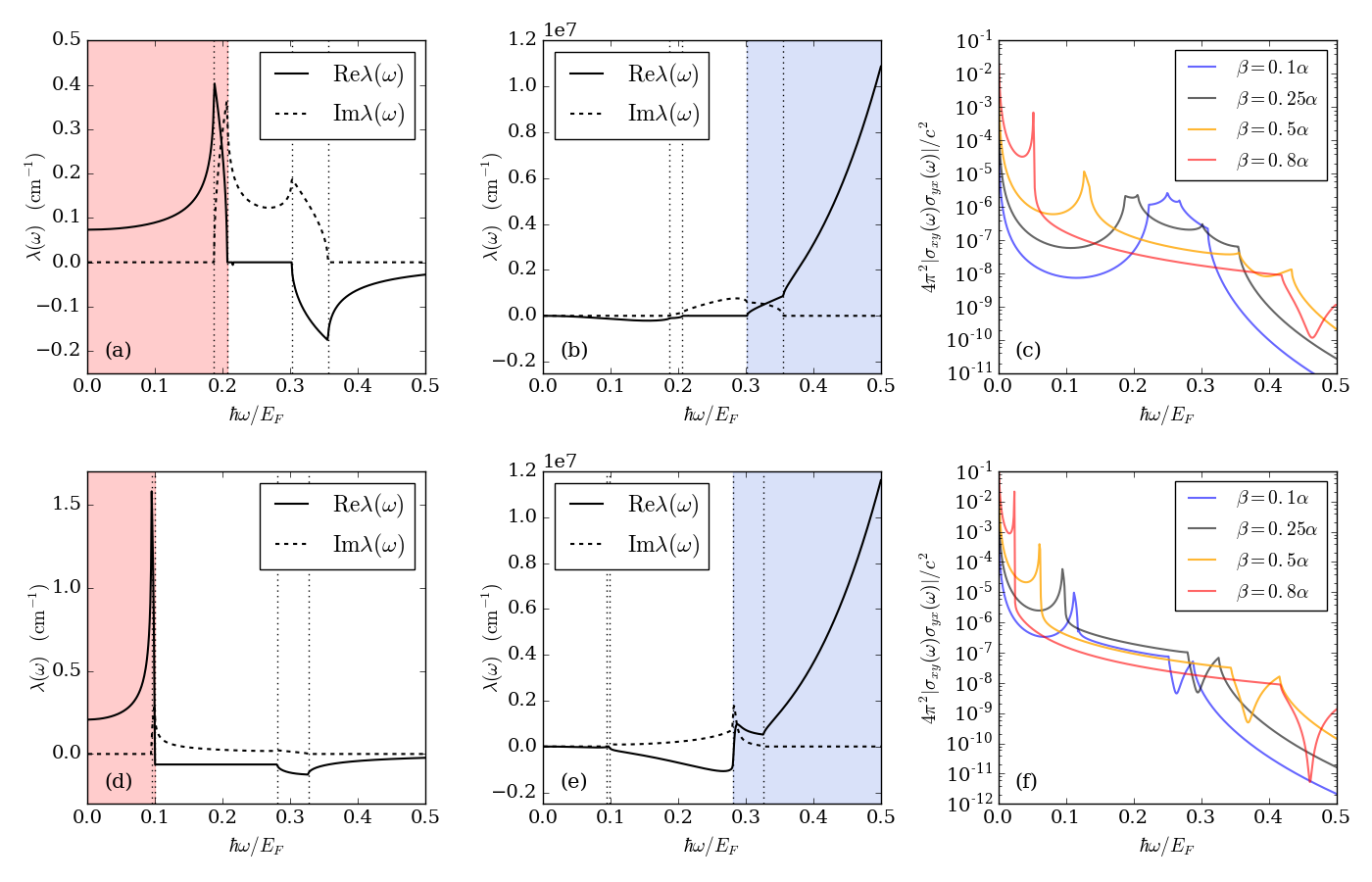}
\caption{
Attenuation constants for (a) quasi-TE and (b) quasi-TM hybrid waves in the plane of a semiconductor quantum well surrounded by dielectric material with permittivity $\varepsilon=1$ with no external pumping. The shaded areas mark regions $\mathrm{Re}\,\lambda>0$ where surface waves localized along the $z=0$ plane exist. The Dresselhaus parameter $\beta=0.25\alpha$. The four vertical lines mark the frequencies, from left to right: $\omega_-$, $\omega_a$, $\omega_b$, and $\omega_+$. The degree of hybridization as a function of frequency at different values of relative spin-orbit-coupling strength $\beta/\alpha$ is shown in (c). (d)--(f) demonstrate how (a)--(c) are modified in the presence of off-resonance pumping with $\gamma=0.8$.
}
\label{fig:wavector}
\end{center}
\end{figure*}

The results of the numerical calculations of the conductivity tensor illustrating (\ref{kubo}) are shown by thick blue solid and dashed lines in Fig.~\ref{fig:numerics}.
The green, orange, and red lines in the figure correspond to the presence of the driving field, which will be discussed in detail in Sec.~\ref{sec:pump}. 
In numerical calculations we have taken the Rashba coupling $\alpha=1.6\times10^{-9}$ eV$\,$cm and the Dresselhaus constant $\beta=0.25\alpha$. For a given concentration of charge carriers $n=5\times10^{11}$ cm$^{-2}$ and an effective mass $m=0.055m_e$ (here, $m_e$ is the mass of an electron), the corresponding Fermi energy of the two-dimensional electron gas $E_0=\pi n\hbar^2/m\approx 21.6$ meV. We fixed these parameters throughout our simulations as they are in line with the experimental studies on 2DEG in InAs-based quantum wells \cite{Giglberger2007,Bastidas2007,Ganichev2014}.
These results for the conductivity tensor can be further used to elaborate on the formation and propagation of surface electromagnetic waves in a spin-orbit-coupled system.

\section{Hybrid Surface Waves}\label{sec:waves}

To investigate the propagation of electromagnetic waves we study a planar structure positioned at $z=0$ between two dielectric slabs with real permittivities $\varepsilon_+$ (at $z>0$) and $\varepsilon_-$ (at $z<0$). Consider a monochromatic plane wave propagating along $x$. In this setting, Maxwell's equations allow solutions in the form of two orthogonal modes representing the TE state with nonzero $(H_x,E_y,H_z)$ and the transverse magnetic (TM) state with $(E_x,H_y,E_z)$, respectively. The general solution of the source-free Maxwell equations with boundary conditions $\mathbf{e}_z\times\left(\mathbf{E}_+-\mathbf{E}_-\right)=0$ and $\mathbf{e}_z\times\left(\mathbf{H}_+-\mathbf{H}_-\right)=\frac{4\pi}{c}\sigma\mathbf{E}_\parallel$ (here, $\mathbf{E}_\parallel$ stands for the components of the electric field parallel to the plane of a quantum well) can be searched in the form of a linear superposition of the TE and TM surface modes \cite{Chiu1974,Iorsh2013,Yudin2015,Kumar2016},
\begin{equation}
\left\lbrace\begin{array}{c}
\mathbf{E}_\pm(x,z,t) \\ \mathbf{H}_\pm(x,z,t)
\end{array}\right\rbrace
=\left\lbrace\begin{array}{c}
\mathbf{E}_\pm \\ \mathbf{H}_\pm
\end{array}\right\rbrace e^{-i\omega t+iqx-\lambda_\pm\vert z\vert},
\label{anzatz}
\end{equation}
where we assume that the surface waves are characterized by the complex propagation constant $q=q(\omega)$  along the $x$ axis (its imaginary part accounts for the damping) and $\lambda_\pm$ are related to $q$ by $\lambda_\pm^2=q^2-\varepsilon_\pm k_0^2$. For convergent solutions  (${\rm Re}\,\lambda_{\pm}>0$) $\lambda_\pm$ have the meaning of attenuation constants. From the ansatz Eq.~(\ref{anzatz}) we get
\begin{equation}
\left(\frac{i\sigma_{xx}}{c}+\frac{\varepsilon_+k_0}{4\pi\lambda_+}+\frac{\varepsilon_-k_0}{4\pi\lambda_-}\right)\left(\frac{\lambda_++\lambda_-}{4\pi k_0}-\frac{i\sigma_{yy}}{c}\right)=\frac{\sigma_{xy}\sigma_{yx}}{c^2},
\label{dispersion}
\end{equation}

\noindent where $k_0=\omega/c$ and $c$ is the speed of light. Equation~(\ref{dispersion}) allows solutions in the form of surface waves localized near the plane $z=0$ with ${\rm Re}\,\lambda_{\pm} >0$. Consider the case $\varepsilon_-=\varepsilon_+\equiv\varepsilon$, where the same dielectric material lies on either side of the semiconductor quantum well. Equation~(\ref{dispersion}) yields the quadratic equation on $\lambda\equiv \lambda_+=\lambda_-$. In the absence of the right-hand side of Eq.~(\ref{dispersion}), the two solutions are just TE and TM modes. The presence of the finite off-diagonal components of the conductivity tensor $\sigma_{xy}(\omega)=\sigma_{yx}(\omega)$ mixes TE and TM modes, thus forming hybridized waves. Due to the small magnitude of the conductivity tensor, $\vert\sigma_{xy}(\omega)\vert/c\ll1$, each of the hybridized modes inherits the properties of either TE or TM modes and thus can be referred to as hybridized quasi-TE and quasi-TM modes, respectively. The attenuation constants for quasi-TE and quasi-TM modes at $\varepsilon=1$ are presented in Figs.~\ref{fig:wavector}(a) and \ref{fig:wavector}(b). The red and blue areas in Figs.~\ref{fig:wavector}(a) and \ref{fig:wavector}(b) mark the regions of frequencies where the hybridized waves take a form of surface waves localized near the $z=0$ plane, that is, when $\mathrm{Re}\,\lambda>0$. 

Because it is hard to perform an exact analytical treatment of the nonlinear dispersion relation (\ref{dispersion}) is in a closed analytical form, we resort to the perturbative approach.
Indeed, the results of our calculations of the optical conductivity shown in Fig. \ref{fig:numerics} reveal the small absolute magnitude of the conductivity tensor. This allows us to elaborate a simple model with $\varepsilon_+=\varepsilon_-\equiv\varepsilon$, making predictions which are quantitatively correct while illustrating the effect of hybridization. 
Using $4\pi\vert\sigma_{ab}(\omega)\vert/c\ll 1$, 
the dispersion relation (\ref{dispersion}) yields the solution
\begin{equation}\label{qte}
\lambda(\omega)\approx i\frac{2\pi\sigma_{yy}(\omega)}{c^2}\left(1+\frac{4\pi^2}{\varepsilon c^2}\sigma_{xy}(\omega)\sigma_{yx}(\omega)\right)
\end{equation}

\noindent for quasi-TE waves, and

\begin{equation}\label{qtm}
\lambda(\omega)\approx i\frac{\varepsilon\omega}{2\pi\sigma_{xx}(\omega)}\left(1-\frac{4\pi^2}{\varepsilon c^2}\sigma_{xy}(\omega)\sigma_{yx}(\omega)\right)
\end{equation}

\noindent for quasi-TM waves, which are in good agreement with
numerically calculated solutions of the full model~\eqref{dispersion}. The terms in parentheses in~\eqref{qte} and ~\eqref{qtm} demonstrate the effect of the off-diagonal terms of the conductivity tensor on the attenuation parameters. Because the hybridization between TE and TM modes is determined by the magnitude of the right-hand side of (\ref{dispersion}), the quantity $4\pi^2 |\sigma_{xy}(\omega)\sigma_{yx}(\omega)|/c^2$ thus quantifies the degree of hybridization. The interplay of the Dresselhaus and Rashba spin-orbit couplings then affects the hybridization of the modes by changing the nonzero diagonal terms of the conductivity tensor. Figure~\ref{fig:wavector}(c) shows how the degree of hybridization is modified when the strength of the Dresselhaus coupling is varied.

\section{Hybrid surface waves under intense pumping}\label{sec:pump} 

The range of frequencies at which the hybrid surface waves considered above exist can be tuned all optically by an external off-resonant coherent pump of frequency $\Omega$, as we will demonstrate below. In the case of a linearly polarized pump, the latter introduces additional anisotropy in the system via renormalization of the parameters of the effective SOI Hamiltonian.  In the high-frequency regime the formal derivation of the effective time-independent Hamiltonian can be performed in a rather intuitive way, and the rigorous mathematical procedure is based on either Floquet-Magnus expansion \cite{Casas2001,Blanes2009} or the Brillouin-Wigner perturbation theory \cite{Mikami2016}: If this is the case, Floquet bands are nearly uncoupled, while the corresponding Floquet Hamiltonian becomes almost block diagonal, which results in a very weak dependence on time of the effective Floquet operator. 
We solve directly the evolution equation that governs the dynamics of Rashba-Dresselhaus spin-orbit-coupled electron gas under an intense linearly polarized field. 
The time dependence is introduced to the Hamiltonian via the electromagnetic vector potential $\mathbf{p}\rightarrow\mathbf{p}-e\mathbf{A}(t)/c$ of the external driving $\mathbf{E}=\mathbf{E}_0\cos\Omega t$ of frequency $\Omega$ and electric field strength $E_0$; the driven field is assumed to be linearly polarized along the $y$ axis, $\mathbf{E}_0=E_0\mathbf{e}_y$. The conducting properties of two-dimensional spin-orbit-coupled electron gas were extensively studied in the past \cite{Schliemann2003a,Maytorena2006}, however, the regime of strong light-matter coupling with an external field has so far been neglected in spite of its particular importance for the design of nanophotonic integrated circuits. The renormalized SOI Hamiltonian can be cast in the following form \cite{Sheremet2016,Yudin2016},

\begin{equation}\label{spinorbit}
\hat H_\mathrm{SO}^\prime=\left(\alpha p_y+\beta^\prime p_x\right)\hat\sigma_x-\left(\alpha^\prime p_x+\beta p_y\right)\hat\sigma_y
\end{equation}

\noindent where the renormalized values

\begin{equation}\label{r}
\alpha^\prime=\alpha\left[1-\frac{\alpha^2-\beta^2}{\alpha^2+\beta^2}\Big(1-J_0(2\gamma)\Big)\right]
\end{equation}

\noindent and 

\begin{equation}\label{d}
\beta^\prime=\beta\left[1+\frac{\alpha^2-\beta^2}{\alpha^2+\beta^2}\Big(1-J_0(2\gamma)\Big)\right],
\end{equation}

\noindent are determined by the light-matter coupling constant

\begin{equation}\label{lm}
\gamma=eE_0\sqrt{\alpha^2+\beta^2}/(\hbar\Omega^2).
\end{equation}

For a driven quantum system the validity of the derived effective Hamiltonian is restricted to the frequency of driving being the dominant energy scale in the system. More specifically,
\begin{equation}\label{criterion}
\left\vert\frac{\left(\alpha^2-\beta^2\right)p_xJ_n^2(2\gamma)/J_0(2\gamma)}{\sqrt{\alpha^2+\beta^2}\left(n\hbar\Omega+2p_y\sqrt{\alpha^2+\beta^2}\right)+4\alpha\beta p_x}\right\vert\ll1,
\end{equation}
\noindent where $J_n(2\gamma)$ denotes the $n$th-order Bessel function of the first kind. The new boundaries of the electron-hole continuum $\hbar\omega_\pm$ can be interpreted as photon maximal and minimal energies required to induce an optical transition between spin-split subbands. The absorption bandwidth $\hbar\delta\omega=\hbar\omega_+-\hbar\omega_-$ therefore has to satisfy $\delta\omega\ll\Omega$ to validate the use of high-frequency expansion. In addition, expression (\ref{criterion}) clearly manifests that $J_0(2\gamma)$ has to be distinct from zero.

Because the effect of off-resonant pumping is to  renormalize the parameters of the Hamiltonian, this also affects the components of the conductivity tensor. The green, orange, and red lines in Fig.~\ref{fig:numerics} represent 
components of the conductivity tensor in the presence of the driving field with the light-matter coupling strengths $\gamma=0.4, 0.6$, and $0.8$, respectively. The effect of the pumping on the hybrid waves is shown in Figs.~\ref{fig:wavector}(d) and \ref{fig:wavector}(e). Note that the  interval of the frequencies where solutions in the form of surface waves exist can be controlled by changing the intensity of the driving field. 
The effects of the driving field on the hybridization between TE and TM modes is demonstrated in Fig.~\ref{fig:wavector}(f), which is to be compared with the case when the driving is absent, Fig.~\ref{fig:wavector}(c). Notice that while the degree of hybridization of quasi-TE waves increases for larger Dresselhaus strengths, it increases even further in the presence of the driving.

\section{Conclusions and outlook}\label{sec:conclusions}

In summary, we have shown that 2D Rashba-Dresselhaus material supports the propagation of hybridized TE-TM surface waves. The latter is closely associated with the appearance of the off-diagonal components of the conductivity tensor due to the simultaneous presence of both Rashba and Dresselhaus SOIs. Although its absolute value is not large enough to lead to dramatic consequences, it induces hybridization between the waves of two different polarizations. We calculated the propagation and attenuation constants and showed how the range of existence of hybridized TE-TM surface waves can be controlled optically by external coherent pumping. We expect that the presented results will trigger experimental activity in this area.

$
$

\section*{Acknowledgments}

We acknowledge support of the Singaporean Ministry of Education under AcRF Tier 2 Grant No. MOE2015-T2-1-055 and the Ministry of Education and Science of the Russian Federation under Increase Competitiveness Program 5-100. D.Y. acknowledges support from RFBR Project No. 16-32-60040. I.A.S. thanks Horizon2020 ITN NOTEDEV and Rannis excellence Grant No. 163082-051.

\end{document}